\documentclass[10pt,a4paper,conference]{IEEEtran}
\usepackage[T1]{fontenc}
\usepackage[latin9]{inputenc}
\usepackage{color}
\usepackage{float}
\usepackage{amsmath}
\usepackage{amsthm}
\usepackage{amssymb}
\usepackage{graphicx}
\usepackage[unicode=true,
 bookmarks=true,bookmarksnumbered=true,bookmarksopen=true,bookmarksopenlevel=1,
 breaklinks=false,pdfborder={0 0 0},pdfborderstyle={},backref=false,colorlinks=false]
 {hyperref}
\hypersetup{pdftitle={Your Title},
 pdfauthor={Your Name},
 pdfborderstyle={},pdfpagelayout=OneColumn,pdfnewwindow=true,pdfstartview=XYZ,plainpages=false}

\makeatletter

\pdfpageheight\paperheight
\pdfpagewidth\paperwidth

\providecommand{\tabularnewline}{\\}
\floatstyle{ruled}
\newfloat{algorithm}{tbp}{loa}
\providecommand{\algorithmname}{Algorithm}
\floatname{algorithm}{\protect\algorithmname}

\theoremstyle{plain}
\newtheorem{thm}{\protect\theoremname}
\theoremstyle{definition}
\newtheorem{defn}[thm]{\protect\definitionname}

\usepackage{epsfig}
\usepackage{amsthm}
\usepackage{algorithm}
\usepackage{algorithmic}

\usepackage{caption}

\newcounter{tempEquationCounter}
\newcounter{thisEquationNumber}
\newenvironment{floatEq}{\setcounter{thisEquationNumber}{\value{equation}}\addtocounter{equation}{1}% record equation as happened and remember number
\begin{figure*}[!t]% float following equation across columns
\normalsize\setcounter{tempEquationCounter}{\value{equation}}% record current equation number in floated location
\setcounter{equation}{\value{thisEquationNumber}}% use previous equation number
}{\setcounter{equation}{\value{tempEquationCounter}}% set back to equation number in floated location
\hrulefill\vspace*{4pt}% add a horizontal rule separator
\end{figure*}% end float environment

}

\providecommand{\definitionname}{Definition}
\providecommand{\theoremname}{Theorem}

\makeatother

\providecommand{\definitionname}{Definition}
\providecommand{\theoremname}{Theorem}

\begin{document}
\vspace{-6.0cm}
\title{Path Selection and Rate Allocation in Self-Backhauled mmWave Networks}\vspace{-2.0cm}

\author{\IEEEauthorblockN{Trung Kien Vu\IEEEauthorrefmark{1},~Chen-Feng~Liu\IEEEauthorrefmark{1},~Mehdi~Bennis\IEEEauthorrefmark{1},~Mérouane
Debbah\IEEEauthorrefmark{2}\IEEEauthorrefmark{3}, and Matti~Latva-aho\IEEEauthorrefmark{1}} \IEEEauthorblockA{\IEEEauthorrefmark{1}Centre for Wireless Communications, University
of Oulu, Finland \\
 \IEEEauthorrefmark{2}Mathematical and Algorithmic Sciences Lab,
Huawei France R$\&$D, Paris, France \\
 \IEEEauthorrefmark{3}CentraleSupelec, Universite Paris-Saclay, Gif-sur-Yvette,
France\\
 E-mail: \{trungkien.vu, chen-feng.liu, mehdi.bennis, matti.latva-aho\}@oulu.fi,
merouane.debbah@huawei.com}}
\maketitle
\begin{abstract}
We investigate the problem of multi-hop scheduling in self-backhauled
millimeter wave (mmWave) networks\footnote{This paper was presented at the IEEE WCNC 2018 Conference, MAC9 -
mmWave MAC Design, in Barcelona, Catalonia, Spain, April 18, 2018.}. Owing to the high path loss and blockage of mmWave links, multi-hop
paths/routes between the macro base station and the intended users
via full-duplex small cells need to be carefully selected. This paper
addresses the fundamental question: ``\textit{how to select the best
paths and how to allocate rates over these paths subject to latency
constraints}?'' To answer these questions, we propose a new system
design, which factors in mmWave-specific channel variations and network
dynamics. The problem is cast as a network utility maximization subject
to a bounded delay constraint and network stability\textit{.} The
studied problem is decoupled into: $\left(i\right)$ a path/route
selection and $\left(ii\right)$ rate allocation, whereby learning
the best paths is done by means of a \textit{reinforcement} \textit{learning}
algorithm, and the rate allocation is solved by applying the successive
convex approximation method. Via numerical results, our approach ensures
reliable communication with a guaranteed probability of $99.9999\%$,
and reduces latency by $50.64\%$ and $92.9\%$ as compared to \textit{baselines}.
\end{abstract}


\IEEEpeerreviewmaketitle{}

\vspace{-0.8em}

\section{Introduction}

The fifth generation (5G) networks are required to support high data
rates of multiple gigabits per second (Gbps) and to have 50 billion
connected devices by 2020 \cite{5GWhat}. In parallel to that, due
to the current scarcity of wireless spectrum, both academia and industry
have paid attention to the underutilized frequency bands ($30$-$300$
GHz) \cite{5GWhat,2013millimeter}. The required capacity increase
can be achieved by $\left(i\right)$ advanced spectral-efficient transmission
techniques, e.g., massive multiple-input multiple-output (MIMO); and
$\left(ii\right)$ ultra-dense self-backhauled small cell deployments~\cite{2015SmallCell,Vu_LB}.
Although mmWave frequency bands offer huge bandwidth, operating at
higher frequency bands experiences high propagation attenuation \cite{2013millimeter},
which requires smart beamforming to achieve highly directional gains.
Owing to the short wavelength, mmWave frequency bands enable packing
a massive number of antennas into highly directional beamforming over
a short distance as compared to the conventional frequency bands~\cite{2013millimeter}.
Besides that, mmWave communication requires higher transmit power
and is very sensitive to blockage, when transmitting over a long distance
\cite{2013millimeter,Vu_LB}. Hence, instead of using a single hop
\cite{Vu_LB,vu2017ultra}, a multi-hop self-backhauling architecture
is a promising solution \cite{2009multihop,2013backhaul}.

Focusing on maximizing the quality of multimedia applications, the
authors in \cite{2016mmW_D2D} studied multi-hop routing for device-to-device
communication. The work \cite{2017multihop} studied the multi-hop
relaying transmission challenges for mmWave systems. Therein, taking
traffic dynamics and link qualities into account, \cite{2016mmW_D2D,2017multihop}
aimed at maximizing the network throughput. In addition, path selection
and multi-path congestion control was studied in \cite{key2011pathselection}
in which the aggregate utility is increased as more paths are provided.

Despite the interesting results of the aforementioned works, using
multi-hop transmissions raises the issue of increased delay which
has been generally ignored. Note that the issues of latency and reliability
are two key components in 5G networks and beyond \cite{2018ultra}.
Moreover, splitting data into too many paths leads to increased signaling
overhead and causes network congestion. Hence, there is a need for
fast and efficient multi-hop multi-path scheduling with respect to
traffic dynamics and channel fluctuations in self-backhauled mmWave
networks. Our previous studies focused on single-hop ultra-reliable
low latency communication (URLLC)-centric transmission in mmWave networks
\cite{vu2017ultra}. In this work, we further extend the previous
work to the multi-hop multi-path wireless backhaul scenario and study
a joint path selection and rate allocation problem. In summary, we
address two fundamental aspects enabling multi-hop multi-path self-backhauled
mmWave networks: $\left(i\right)$ how to select the best paths while
taking traffic dynamics and link qualities into account; $\left(ii\right)$
how to capture elements of URLLC while maximizing the network utility.

\subsection{Main contribution}

Considering a multi-hop multi-path self-backhauled mmWave network,
we propose an efficient system design to support URLLC. In particular,
our goal is to maximize a general network utility subject to network
stability and the delay bound violation constraint with a tolerable
probability (reliability). Leveraging Lyapunov stochastic optimization
\cite{neely2010S}, the studied problem is decoupled into multi-hop
path/route selection and rate allocation sub-problems. The challenging
questions we seek to address are: $(i)$ \textit{over which paths
should the traffic flow be forwarded}? and $(ii)$ \textit{what is
the data rate per flow/sub-flow} while ensuring low-latency and ultra-reliability
constraints? To answer these questions, we utilize \textit{regret}
\textit{learning} techniques to exploit the benefits of the historical
information which aids in selecting the best paths. For rate allocation,
the corresponding mathematical problem belongs to a non-convex combinatorial
program \cite{2004convex}. By exploiting the hidden convexity of
the problem, we propose an iterative rate allocation algorithm based
on the second-order cone program (SOCP) to obtain a local optimal
of the approximated convex problem. Numerical results verify the effectiveness
of the proposed path selection and rate allocation solution.

\section{System Model}

\label{SystemModel}

\begin{figure}
\includegraphics[width=0.5\textwidth]{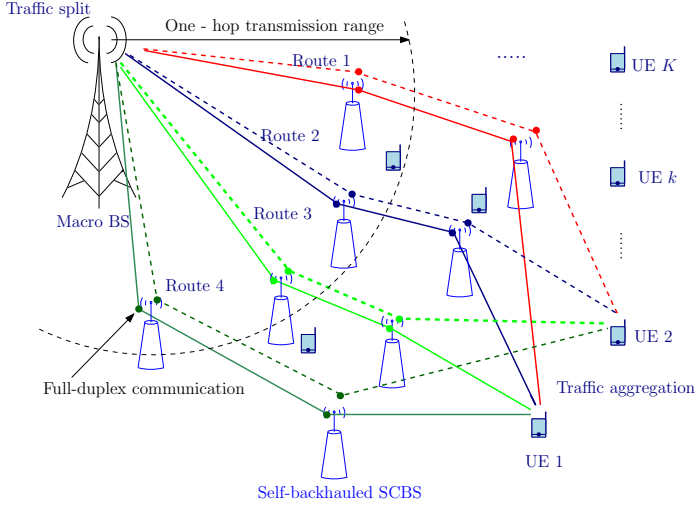}
\caption{Illustration of 5G multi-hop self-backhauled mmWave networks.}
\label{Example-1}
\end{figure}

Let us consider a downlink (DL) transmission of a multi-hop heterogeneous
cellular network (HCN) which consists of a macro base station (MBS),
a set of $B$ self-backhauled small cell base stations (SCBSs), and
a set ${\cal K}$ of $K$ single-antenna user equipments (UEs) as
shown in Fig \ref{Example-1}. Let ${\cal B}=\{0,1,\cdots,B\}$ denote
the set of all base stations (BSs) in which index ${\rm 0}$ refers
to the MBS. The in-band wireless backhaul is used to provide backhaul
among BSs \cite{Vu2016}. A full-duplex (FD) transmission protocol
is assumed at SCBS capable with perfect self-interference cancellation
(SIC) capabilities. Each BS is equipped with $N_{b}$ transmitting
antennas and we denote the propagation channel between BS $b$ and
UE $k$ as $\mathbf{h}_{(b,k)}=\sqrt{N_{b}}\mathbf{\Theta}_{(b,k)}^{1/2}\mathbf{{\bf w}}_{\left(b,k\right)}$
\cite{Vu_LB}, where $\mathbf{\Theta}_{\left(b,k\right)}\in\mathbb{C}^{N_{b}\times N_{b}}$
depicts the antenna spatial correlation, and the elements of $\mathbf{{\bf w}}_{\left(b,k\right)}\in\mathbb{C}^{N_{b}\times1}$
are independent and identically distributed (i.i.d.) with zero mean
and variance $1/N_{b}$.

The network topology is modeled as a directed graph $\mathcal{G}=(\mathcal{N},\,\mathcal{L})$,
where ${\cal N}=\mathcal{B}\,\cup\mathcal{K}$ represents the set
of nodes including BSs and UEs. ${\cal L}=\{(i,j)|i\in\mathcal{B},j\in\mathcal{N}\}$
denotes the set of all directional edges $(i,j)$ in which nodes $i$
and $j$ are the transmitter and the receiver, respectively.

\begin{table}
\caption{Notations for system model.}
\begin{tabular}{|c|l|}
\hline
Notations & Descriptions\tabularnewline
\hline
\hline
${\cal B}$,${\cal \,{\cal K}}$ & Sets of $\left(B+1\right)$ base stations, $K$ user equipments\tabularnewline
\hline
${\cal {\cal {\cal N}=\mathcal{B}\,\cup\mathcal{K}}}$ & Set of nodes including BSs and UEs\tabularnewline
\hline
${\cal L}$ & Set of all directional edges $(i,j)|i\in\mathcal{B},j\in\mathcal{N}$\tabularnewline
\hline
${\cal F}$ & Set of $F$ flows\tabularnewline
\hline
$\mathcal{Z}_{f}$ & Set of $Z_{f}$ disjoint paths observed by flow $f$\tabularnewline
\hline
$\mathcal{N}_{i}^{({\rm o)}}$  & Set of the next hops from node $i$\tabularnewline
\hline
$i_{f}^{({\rm I)}}$ & Previous hop of flow $f$ to BS $i$ \tabularnewline
\hline
$i_{f}^{({\rm o)}}$ & Next hop of flow $f$ from BS $i$ \tabularnewline
\hline
$p_{(i,j)}^{f}$ & Transmit power of node $i$ to node $j$ for flow $f$\tabularnewline
\hline
$z_{f}^{m}=1$ & Path $m$ is used to send data for flow $f$\tabularnewline
\hline
$\pi_{f}^{m}$ & Probability of choosing path $m$ for flow $f$\tabularnewline
\hline
\end{tabular}\label{notations1}
\end{table}

We consider a queuing network operating in discrete time $t\in\mathbb{Z}^{+}$.
There are $F$ independent data at the MBS. Each data traffic is destined
for only one UE, whereas one UE can receive multiple data streams,
i.e., $F\geq K$. Hereafter, we refer to data traffic as data flow.
We use $\mathcal{F}$ to represent the set of $F$ data flows/sub-flows.
The MBS can split each flow $f\in{\cal F}$ into multiple sub-flows
which are sent through a set of disjoint paths. The traffic aggregation
capability is assumed at the UEs~\cite{2013dualconnect}.

We assume that there exits $Z_{f}$ number of disjoint paths from
the MBS to the UE for flow $f$. For any disjoint path $m\in\left\{ 1,\cdots,Z_{f}\right\} $,
we denote $\mathcal{Z}_{f}^{m}$ as the path state, which contains
all path information such as topology and queue states for every hop.
Let $\mathcal{Z}_{f}=\{\mathcal{Z}_{f}^{1},\cdots,\mathcal{Z}_{f}^{m},\cdots,\mathcal{Z}_{f}^{Z_{f}}\}$
denote path states observed by flow $f$. We use the flow-split indicator
vector $\mathbf{z}_{f}=\left(z_{f}^{1},\cdots,\,z_{f}^{Z_{f}}\right)$
to denote how the MBS splits flow $f$, where $z_{f}^{m}=1$ means
path $m$ is used to send data for flow $f$; otherwise, $z_{f}^{m}=0$.
Let $\mathcal{N}_{i}^{({\rm o)}}$ denote the set of the next hops
from node $i$ via a directional edge. We denote the next hop and
the previous hop of flow $f\,$ from and to BS $i$ as $i_{f}^{({\rm o)}}$and
$i_{f}^{({\rm I)}}$, respectively. Table \ref{notations1} shows
the notations, which used through this paper.

In addition, ${\bf h}=\big({\bf h}_{(i,j)}|(i,j)\in\mathcal{L}\big)$
is the channel propagation vector, and we denote $p_{(i,j)}^{f}$
as the transmit power of node $i$ assigned to node $j$ for flow
$f$, such that $\sum_{f\in F}\sum_{j\in{\cal N}_{i}^{{\rm (o)}}}p_{(i,j)}^{f}\leq P_{i}^{{\rm max}},$
where $P_{i}^{{\rm max}}$ is the maximum transmit power of node $i$.
We have the power constraint as

\vspace{-1em}
\begin{align}
\!\mathcal{P} & =\bigg\{ p_{(i,j)}^{f}\geq0,i,j\in\mathcal{N},\Big|\sum_{f\in{\cal F}}\sum_{j\in{\cal N}_{i}^{{\rm (o)}}}p_{(i,j)}^{f}\leq P_{i}^{{\rm max}}\bigg\}.\label{eq:powerconstraint-0}
\end{align}
Vector $\mathbf{p}=(p_{(i,j)}^{f}|\forall i,j\in{\cal N},\forall f\in{\cal F})$
denotes the transmit power over all flows.

Here, we assume that each BS adopts the hybrid beamforming architecture,
which enjoys both analog and digital beamforming techniques \cite{2017hybrid}.
For the analog beamforming, let $g_{(i,j)}^{(t)}$ and $g_{(i,j)}^{(r)}$
denote the transmitter and receiver beamforming gain at the transmitter
$i$ and the receiver $j$, respectively. In addition, we use $\omega_{(i,j)}^{(t)}$
and $\omega_{(i,j)}^{(r)}$ to represent the angles deviating from
the strongest path between the transmitter $i$ and the receiver $j$.
Also, let $\theta_{(i,j)}^{(t)}$ and $\theta_{(i,j)}^{(r)}$ denote
the beamwidth at the transmitter $i$ and the receiver $j$, respectively.
We denote $\boldsymbol{\theta}$ as a vector of the transmitter beamwidth
of all BSs. We adapt the widely used antenna radiation pattern model
\cite{2017hybrid,2014beamwidth} to determine the beamforming gain
as

\begin{align*}
g_{(i,j)}\left(\omega_{(i,j)},\theta_{(i,j)}\right) & =\begin{cases}
\frac{2\pi-\left(2\pi-\theta_{(i,j)}\right)\eta}{\theta_{(i,j)}}, & \text{if}\:|\omega_{(i,j)}|\leq\frac{\theta_{(i,j)}}{2},\\
\eta, & \text{otherwise,}
\end{cases}
\end{align*}
where $0<\eta\ll1$ is the side lobe gain. \textcolor{black}{For the
digital beamforming phase, we apply the linear precoding scheme ${\bf v}_{(i,j)}$,
i.e., for the conjugate precoding, ${\bf v}({\bf h}_{(i,j)})=\hat{{\bf h}}_{(i,j)}$.
Here, $\hat{{\bf h}}_{(i,j)}$ is the estimated channel of ${\bf h}_{(i,j)}$,
such that
\[
\textcolor{blue}{ \hat{{\bf h}}_{(i,j)}= \sqrt{N_{i}}\Theta_{(i,j)}^{1/2} \left(\sqrt{1-\tau_{j}^{2}}{\bf w}_{(i,j)}+\tau_{j}\hat{{\bf w}}_{(i,j)}\right) },
\]
where $\tau_{j}\in[0,1]$ reflects the estimation accuracy for receiver
$j$, if $\tau_{j}=0$, the perfect channel state information is assumed
at the transmitters \cite{2006fastCSI}. $\hat{{\bf w}}_{(i,j)}\in\mathbb{C}^{N_{i}}$
is the estimated noise vector, also modeled as a random matrix with
zero mean and variance of $\frac{1}{N_{i}}$ \cite{Vu_LB}. Based
on the hybrid model \cite{2017hybrid}, the Ergodic achievable rate}\footnote{\textcolor{black}{Note that we omit the beam search/track time, since
it can be done fast and is very small as compared the transmission
time \cite{2017tracking}. }}\textcolor{black}{{} of and the receiver $j$ from the transmitter $i$
can be calculated as per \eqref{rate_calculate}, where $p_{(i,j)}$
and $p_{(i^{\prime},j)}$ are the transmit power from the transmitter
$i$ and $i^{\prime}$ to the receiver $j$, respectively, and the
thermal noise of receiver $j$ is $\eta_{(i,j)}$. In addition, $\text{W}$
denotes the system bandwidth of the mmWave frequency band.}

\begin{floatEq}
\begin{align}
R^{\left(i,j\right)} & =\mathbb{E}_{{\bf h},\,{\bf p}}\left[\text{W}\log_{2}\left(1+\frac{p_{(i,j)}g_{(i,j)}^{(t)}g_{(i,j)}^{(r)}|{\bf h}_{(i,j)}^{\dagger}{\bf v}_{(i,j)}|^{2}}{\sum_{i^{\prime}\neq i}p_{(i^{\prime},j)}g_{(i^{\prime},j)}^{(t)}g_{(i^{\prime},j)}^{(r)}|{\bf h}_{(i^{\prime},j)}^{\dagger}{\bf v}_{(i^{\prime},j)}|^{2}+\eta_{(i,j)}}\right)\right]\label{rate_calculate}
\end{align}
\end{floatEq}

Therefore, for a given channel state and transmit power, the data
rate in edge $(i,j)$ over flow $f$ can be posted as a function of
channel state and transmit power, i.e., $R_{f}^{(i,j)}\left({\bf h},{\bf \,p}\right)$,
such that $\sum_{f\in{\cal F}}R_{f}^{\left(i,j\right)}=R^{\left(i,j\right)}$.
We denote $\mathbf{R}=(R_{f}^{\left(i,j\right)}|\forall i,j\in{\cal N},\forall f\in{\cal F})$
as a vector of data rates over all flows.

Let $Q_{f}^{i}(t)$ denote the queue length at BS $i$ at time slot
$t$ for flow $f$. The queue length evolution at the MBS $i=0$ is
\vspace{-0.5em}
\begin{equation}
Q_{f}^{i}(t+1)=\bigg[Q_{f}^{i}(t)-\sum\limits _{m=1,i_{f}^{({\rm o)}}\in\mathcal{N}_{i}^{({\rm o)}}}^{Z_{f}}R_{f}^{(i,i_{f}^{({\rm o)}})}(t),\:0\bigg]^{+}+\mu^{f}(t),\label{eq:queue1}
\end{equation}
where $\mu^{f}(t)$ is the data arrival at the MBS during slot $t$,
which is independent and identical distributed (i.i.d.) over time
with a mean value $\bar{\mu}^{f}$. Due to the disjoint paths, for
each flow $f\,$ the incoming rate from the previous hop $i_{f}^{({\rm I)}}$
at the SCBS $i$ is either from another SCBS or the MBS, and thus,
the queue evolution at the SCBS $i=\left\{ 1,\cdots,\,B\right\} $
is given by
\begin{equation}
Q_{f}^{i}(t+1)\leq\bigg[Q_{f}^{i}(t)-R_{f}^{(i,i_{f}^{({\rm o)}})}(t),\:0\bigg]^{+}+R_{f}^{(i_{f}^{({\rm I)}},i)}(t).\label{eq:queue2}
\end{equation}

\begin{defn}
For any vector $\mathbf{x}\left(t\right)=\left(x_{1}\left(t\right),...,x_{\emph{K}}\left(t\right)\right)$,
let $\bar{\mathbf{x}}=\left(\bar{x}_{1},\cdots,\bar{x}_{\emph{K}}\right)$
denote the time average expectation of $\mathbf{x}\left(t\right)$,
where ${\textstyle \bar{\mathbf{x}}\triangleq\lim_{t\to\infty}\frac{1}{t}\sum_{\tau=0}^{t-1}\mathbb{E}\left[\mathbf{x}\left(\tau\right)\right]}$.
\end{defn}
\begin{defn}
For any discrete queue $Q\left(t\right)$ over time slots $t\in\left\{ 0,1,\ldots\right\} $
and $Q\left(t\right)\in R_{+}$,
\end{defn}
\begin{itemize}
\item $Q\left(t\right)$ is strongly stable if ${\textstyle \lim_{t\to\infty}\sup\frac{1}{t}\sum_{\tau=0}^{t-1}\mathbb{E}\left[|Q\left(\tau\right)|\right]<\infty}$.
\item $Q\left(t\right)$ is mean rate stable if ${\textstyle \lim_{t\to\infty}\frac{\mathbb{E}\left[|Q\left(t\right)|\right]}{t}=0}$.
\end{itemize}
A queue network is stable if each queue is stable.

\section{Problem Formulation}

\label{Pro-Form}

Assume that the MBS determines paths to split data flow $f$ with
a given probability distribution, i.e., $\boldsymbol{\pi}_{f}=\big(\pi_{f}^{1},\cdots,\pi_{f}^{Z_{f}}\big)$,
where for each $m\in{\cal Z}_{f}$ we have $\pi_{f}^{m}=\text{Pr}\left(z_{f}=z_{f}^{m}\right)$.
Here, $\boldsymbol{\pi}_{f}$ is the probability mass function (PMF)
of the flow-split vector, i.e., $\sum_{m=1}^{Z_{f}}\text{Pr}\left(z_{f}^{m}\right)=1$.
We denote $\boldsymbol{\pi}=\left\{ \boldsymbol{\pi}_{1},\cdots,\boldsymbol{\pi}_{f},\cdots,\boldsymbol{\pi}_{F}\right\} \in\Pi$
as the global probability distribution of all flow-split vectors in
which $\Pi$ is the set of all possible global PMFs. Let $\bar{x}_{f}$
denote the achievable average rate of flow $f$, where $\bar{x}_{f}\triangleq\lim\limits _{t\to\infty}\frac{1}{t}\sum_{\tau=0}^{t-1}x_{f}\left(\tau\right)$
and $x_{f}\left(\tau\right)=\sum_{m=1,i_{f}^{({\rm o)}}\in\mathcal{N}_{i}^{({\rm o)}}}^{Z_{f}}\mathbb{E}_{{\bf h},{\bf p}}\big[\pi_{f}^{m}R_{f}^{(i,i_{f}^{({\rm o)}})}(\tau)\big]\Big|_{i=0}$.
We assume that the achievable rate is bounded, i.e.,

\vspace{-1em}
\begin{eqnarray}
0 & \leq & x_{f}(t)\leq a_{f}^{\max},\label{ratebounds}
\end{eqnarray}
where $a_{f}^{\max}$ is the maximum achievable rate of flow $f$
at every time $t$. Vector $\bar{{\bf x}}=\left(\bar{x}_{1},\cdots,\bar{x}_{F}\right)$
denotes the time average of rates over all flows. Let $\mathcal{R}$
denote the rate region, which is defined as the convex hull of the
average rates, i.e., $\bar{{\bf x}}\in\mathcal{R}$.

We define $U_{0}$ as a network utility function, i.e., $U_{0}\left(\bar{{\bf x}}\right)=\sum_{f\in{\cal F}}U\left(\bar{x}_{f}\right)$.
Here, $U(\cdot)$ is assumed to be a twice differentiable, concave,
and increasing $L\text{-Lipschitz}$ function for all $\bar{{\bf x}}\geq0$.
According to Little's law \cite{2008little}, the queuing delay is
defined as the ratio of the queue length to the average arrival rate.
By taking into account the probabilistic delay constraints for each
flow/subflow, the following network utility maximization (NUM) is
formulated as:

\vspace{-1em}\begin{subequations}\label{OP1}
\begin{align}
\hspace{-1em}\mbox{OP:}\:\max_{\boldsymbol{\pi},\mathbf{\,x},\,{\bf p}} & \quad U_{0}(\bar{\mathbf{x}})\label{eq:obj1}\\
\text{subject to} & \text{\quad Pr\ensuremath{\left(\frac{\text{\ensuremath{Q_{f}^{i}(t)}}}{\bar{\mu}_{f}}\geq\beta\right)\leq\epsilon}},\forall t,f\in\mathcal{F},i\in\mathcal{B},\label{eq:delayconst}\\
 & \quad\lim_{t\rightarrow\infty}\frac{\mathbb{E}\left[|Q_{f}^{i}|\right]}{t}=0,\forall f\in\mathcal{F},\forall i\in\mathcal{B},\label{eq:queueStability}\\
 & \mathbf{\quad\mathbf{x}}(t)\in\mathcal{R},\label{Rateconst}\\
 & \quad\boldsymbol{\pi}\in\Pi,\label{Piconst}\\
 & \quad\text{and }\:\eqref{eq:powerconstraint-0},\:\eqref{ratebounds},\nonumber
\end{align}
\end{subequations}where $\text{Pr}(\cdot)$ denotes the probability
operator, $\beta$ reflects the maximum allowed delay requirement
for UEs, and $\epsilon\ll1$ is the target probability for reliable
communication. The probabilistic delay constraint \eqref{eq:delayconst}
implies that the probability that the delay for each flow at node
$b$ is greater than $\beta$ is very small, which captures the constraints
of ultra-low latency and reliable communication. It is also used to
avoid congestion for each flow $f$ at any point (BS) in the network,
if the queue length is greater than $\beta\bar{\mu}^{f}$. More importantly,
\eqref{eq:delayconst} forces the transmission of all BSs, and~\eqref{eq:queueStability}
maintains network stability.

The above problem has a non-linear probabilistic constraint \eqref{eq:delayconst},
which cannot be solved directly. Hence, we replace the non-linear
constraint \eqref{eq:delayconst} with a linear deterministic equivalent
by applying Markov's inequality \cite{2013queue,vu2017ultra} such
that $\text{Pr}\left(X\geq a\right)\leq\mathbb{E}\left[X\right]/a$
for a non-negative random variable $X$ and $a>0$. Thus, we relax
\eqref{eq:delayconst} as
\begin{equation}
\mathbb{E}\big[Q_{f}^{i}(t)\big]\leq\bar{\mu}^{f}\epsilon\beta.\label{appro_delayconst}
\end{equation}
Assuming that $\mu^{f}(t)$ follows a Poisson arrival process~\cite{2013queue},
we derive the expected queue length in \eqref{eq:queue1} for $i=0$
as
\begin{align}
\mathbb{E}[Q_{f}^{i}(t)] & =t\bar{\mu}^{f}-\sum_{\tau=1}^{t}\sum_{m=1,i_{f}^{({\rm o)}}\in\mathcal{N}_{i}^{({\rm o)}}}\pi_{f}^{m}R_{f}^{(i,i_{f}^{({\rm o)}})}(\tau),\label{queue1-1}
\end{align}
and the expected queue length in \eqref{eq:queue2}, for each SCBS,
i.e.,
\begin{equation}
\mathbb{E}[Q_{f}^{i}(t)]=\sum_{\tau=1}^{t}\sum_{m}\pi_{f}^{m}\bigg(R_{f}^{(i_{f}^{({\rm I)}},i)}\left(\tau\right)-R_{f}^{(i,i_{f}^{({\rm o)}})}(\tau)\bigg).\label{queue2-1}
\end{equation}
Subsequently, combining the constraints \eqref{appro_delayconst}
and \eqref{queue1-1}, we obtain, for MBS $i=0$,
\begin{align}
 & \bar{\mu}^{f}(t-\epsilon\beta)-\sum_{\tau=1}^{t-1}\sum_{m=1,i_{f}^{({\rm o)}}\in\mathcal{N}_{i}^{({\rm o)}}}\pi_{f}^{m}R_{f}^{(i,i_{f}^{({\rm o)}})}(\tau)\nonumber \\
 & \leq\sum_{m=1,i_{f}^{({\rm o)}}\in\mathcal{N}_{i}^{({\rm o)}}}\pi_{f}^{m}R_{f}^{(i,i_{f}^{({\rm o)}})}\left(t\right).\label{eq:delayconst01}
\end{align}
Similarly, for each SCBS $i=\{1,\cdots,B\}$, we have
\begin{align}
 & -\bar{\mu}^{f}\epsilon\beta+\sum_{\tau=1}^{t-1}\sum_{m}\pi_{f}^{m}\bigg(R_{f}^{(i_{f}^{({\rm I)}},i)}\left(\tau\right)-R_{f}^{(i,i_{f}^{({\rm o)}})}\left(\tau\right)\bigg)\nonumber \\
 & \leq\sum_{m}\pi_{f}^{m}\bigg(R_{f}^{(i,i_{f}^{({\rm o)}})}\left(t\right)-R_{f}^{(i_{f}^{({\rm I)}},i)}\left(t\right)\bigg),\label{eq:delayconst02}
\end{align}
by combining \eqref{appro_delayconst} and \eqref{queue2-1}. With
the aid of the above derivations, we consider \eqref{eq:delayconst01}
and \eqref{eq:delayconst02} instead of \eqref{eq:delayconst} in
the original problem \eqref{OP1}. In practice, the statistical information
of all candidate paths to decide $\boldsymbol{\pi}_{f},\forall\,f\in\mathcal{F}$,
is not available beforehand, and thus solving \eqref{OP1} is very
difficult. One solution is that paths are randomly assigned to each
flow which does not guarantee optimality, whereas applying an exhaustive
search is not practical. Therefore, in this work, we propose a low-complexity
approach by invoking tools from Lyapunov stochastic optimization which
achieves the optimal performance without requiring the statistical
information beforehand.

\section{\textcolor{blue}{Proposed Algorithm}}

\label{Design}

In this section, we propose a Lyapunov optimization based framework
in order to solve \eqref{OP1}. To do that, we first introduction
the auxiliary variables to refine the original problem \eqref{OP1}.
Next, we convert the constraints into virtual queues and write the
conditional Lyapunov drift function. Finally, the solution of equivalent
problem is obtained by minimizing the Lyapunov drift and the penalty
from the objective function.

Let us start by rewriting \eqref{OP1} equivalently as \cite{2015receding}

\vspace{-1.5em}\begin{subequations}\label{LP2}
\begin{eqnarray}
\ \mbox{RP:}\,\max_{\bar{\boldsymbol{\varphi}},\boldsymbol{\pi},{\bf p}} &  & U_{0}\left(\bar{\boldsymbol{\varphi}}\right)\label{LP1:obj}\\
\text{subject to} &  & \bar{\varphi_{f}}-\bar{x_{f}}\leq0,\ \forall f\in{\cal F},\label{lp1:setconstraint}\\
 &  & \eqref{eq:powerconstraint-0},\:\eqref{ratebounds},\,\eqref{eq:queueStability},\,\eqref{Piconst},\,\eqref{eq:delayconst01},\,\eqref{eq:delayconst02},\nonumber
\end{eqnarray}
\end{subequations}where the new constraint \eqref{lp1:setconstraint}
is introduced to replace the rate constraint \eqref{Rateconst} with
new auxiliary variables $\boldsymbol{\varphi}=\left(\varphi_{1},\:\cdots,\varphi_{F}\right)$.
In \eqref{lp1:setconstraint}, $\bar{\boldsymbol{\varphi}}\triangleq\lim\limits _{t\to\infty}\frac{1}{t}\sum_{\tau=0}^{t-1}\mathbb{E}\left[|\boldsymbol{\varphi}(\tau)|\right]$.
In order to ensure the inequality constraint \eqref{lp1:setconstraint},
we introduce a virtual queue vector $Y_{f}\left(t\right),$ which
is given by

\vspace{-1em}
\begin{equation}
Y_{f}\left(t+1\right)=\left[Y_{f}\left(t\right)+\varphi_{f}\left(t\right)-x_{f}\left(t\right)\right]^{+},\:\forall f\in{\cal F}.\label{eq:virtualQ}
\end{equation}
Let ${\bf \boldsymbol{\Sigma}}(t)=({\bf Q}(t),\,{\bf Y}(t))$ denote
the queue backlogs, we first write the conditional Lyapunov drift
for slot $t$ as

\begin{equation}
\mathbf{\Delta}({\bf \boldsymbol{\Sigma}}(t))=\mathbb{E}\left[L\left({\bf \boldsymbol{\Sigma}}(t+1)\right)-L\left({\bf \boldsymbol{\Sigma}}(t)\right)|{\bf \boldsymbol{\Sigma}}(t)\right],\label{eq:drift}
\end{equation}
where $L({\bf \boldsymbol{\Sigma}}(t))\triangleq\frac{1}{2}\left[\sum_{f=1}^{F}\sum_{i=0}^{B}Q_{f}^{i}(t)^{2}+\sum_{f=1}^{F}Y_{f}(t)^{2}\right]$
is the quadratic Lyapunov function of ${\bf \boldsymbol{\Sigma}}(t)$
\cite{neely2010S}. We then apply the Lyapunov drift-plus-penalty
technique \cite{neely2010S,2015receding,Vu_LB}, where the solution
of \eqref{LP2} is obtained by minimizing the Lyapunov drift and a
penalty from the objective function, i.e.,

\vspace{-1em}
\begin{eqnarray}
\min &  & \mathbf{\Delta}(\mathbf{{\bf \boldsymbol{\Sigma}}}(t))-\nu\mathbb{E}\left[U_{0}\left(\bar{\boldsymbol{\varphi}}\right)|{\bf \boldsymbol{\Sigma}}(t)\right].\label{eq:driftplusPenalty}
\end{eqnarray}
Here, $\nu$ is a control parameter to trade off utility optimality
and queue length \cite{Vu_LB}. Note that the stability of ${\bf \boldsymbol{\Sigma}}(t)$
assures that the constraints of problem \eqref{eq:queueStability}
and \eqref{lp1:setconstraint} are held. Subsequently, following the
straightforward calculations of the Lyapunov optimization which are
omitted here for space, assuming that $\boldsymbol{\varphi}\in{\cal R}$
and a feasible ${\bf \pi}$ and all possible ${\bf \boldsymbol{\Sigma}}(t)$
for all $t$, we obtain

\vspace{-1em}
\begin{eqnarray}
\eqref{eq:driftplusPenalty} & \leq & \sum_{f=1}^{F}\sum_{i=1}^{B}Q_{f}^{i}\,\mathbb{E}\left[\sum_{m}\pi_{f}^{m}\left(R_{(i_{f}^{({\rm I)}},i)}^{f}-R_{(i,i_{f}^{({\rm o)}})}^{f}\right)|{\bf \boldsymbol{\Sigma}}(t)\right]\nonumber \\
 &  & -\sum_{f=1}^{F}Q_{f}^{i|i=0}\:\mathbb{E}\left[\sum_{m=1,i_{f}^{({\rm o)}}\in\mathcal{N}_{i}^{({\rm o)}}}\pi_{f}^{m}R_{f}^{(i,i_{f}^{({\rm o)}})}|{\bf \boldsymbol{\Sigma}}(t)\right]\label{eq:drift2}\\
 &  & +\sum_{f=1}^{F}\mathbb{E}\left[Y_{f}\varphi_{f}-\nu U\left(\varphi_{f}\right)-Y_{f}x_{f}|{\bf \boldsymbol{\Sigma}}(t)\right]+\Psi.\nonumber
\end{eqnarray}
\textcolor{black}{Here, the constant value $\Psi$ does not influence
the system performance \cite{neely2010S,Vu_LB}. }The solution to
\eqref{LP2} can be obtained by minimizing the upper bound in \eqref{eq:drift2}.
For every slot $t,$ we observe ${\bf \boldsymbol{\Sigma}}(t)$ and
have three decoupled subproblems as follows: The flow-split vector
and the probability distribution are determined by

\vspace{-1em}
\begin{align*}
\mbox{SP1}:\,\min_{\boldsymbol{\pi}} & \qquad\sum_{f=1}^{F}\Xi_{f}\\
\text{subject to} & \qquad\eqref{Piconst},
\end{align*}
where

\vspace{-2em}
\begin{eqnarray*}
\Xi_{f} & = & \sum_{i=1}^{B}Q_{f}^{i}\sum_{m}\pi_{f}^{m}\left(R_{f}^{(i_{f}^{({\rm I)}},i)}-R_{f}^{(i,i_{f}^{({\rm o)}})}\right)\\
 &  & -Q_{f}^{i|i=0}\sum_{m=1,i_{f}^{({\rm o)}}\in\mathcal{N}_{i}^{({\rm o)}}}\pi_{f}^{m}R_{(i,i_{f}^{({\rm o)}})}^{f}.
\end{eqnarray*}
Then, we select the optimal auxiliary variables by solving the following
convex optimization problem

\vspace{-1em}
\begin{eqnarray*}
\mbox{SP2:}\:\min_{\boldsymbol{\varphi}|\boldsymbol{\pi}} &  & \sum_{f=1}^{F}\left[Y_{f}\,\varphi_{f}-\nu U\left(\varphi_{f}\right)\right]\\
\mbox{\mbox{subject to}} &  & \varphi_{f}(t)\geq0,\:\forall f\in{\cal F}.
\end{eqnarray*}
Let $\varphi_{f}^{\ast}$ be the optimal solution obtained by the
first order derivative of the objective function of SP2. Assuming
a logarithmic utility function, we have $\varphi_{f}^{\ast}(t)=\max\left\{ \frac{\nu}{Y_{f}},\:0\right\} .$
Finally, the rate allocation is done by assigning transmit power,
which is obtained by

\vspace{-1em}
\begin{eqnarray*}
\mbox{SP3:}\:\min_{{\bf x},{\bf p}|\boldsymbol{\pi}} &  & \sum_{f=1}^{F}-Y_{f}\,x_{f}\\
\text{subject to} &  & \eqref{eq:powerconstraint-0},\:\eqref{ratebounds},\,\eqref{eq:delayconst01},\,\eqref{eq:delayconst02}.
\end{eqnarray*}

\subsection{\textcolor{blue}{Path Selection}}

Now \textcolor{black}{we leverage regret learning which exploits the
historical system information such as queue state and channel state
to select the optimal paths in SP1 \cite{bennislearning}. The intuition
behind this approach is that the} regret learning method results in
maximizing the long-term utility for each flow. Recall that ${\bf z}_{f}\,$
represents the flow-split vector given to flow $f\,$ and $z_{f}^{m}=1$
means path $m$ is used to send data for flow $f$. The MBS selects
paths for each flow with a given probability (mixed strategy). The
optimal strategies mean that the MBS does not wish to change its strategy
for any flow where any deviation does not offer better utility gain
for all flows. We denote $u_{f}^{m}=u_{f}\left(z_{f}^{m},{\bf z}_{f}^{-m}\right)$
as a utility function of flow $f$ when using path $m$. The vector
${\bf z}_{f}^{-m}$ denotes the flow-split vector excluding path $m$.
The MBS can choose more than one path to deliver data, from $\mbox{SP1}$,
the utility gain of flow $f$ is

\[
u_{f}=\sum_{m}u_{f}^{m}=-\Xi_{f}.
\]
To\textcolor{black}{{} exploit the historical information}, the MBS
determines a flow-split vector for each flow $f$ from ${\cal Z}_{f}$
based on the PMF from the previous stage $t-1$, i.e.,\vspace{-1em}

\begin{equation}
{\bf \boldsymbol{\pi}}_{f}\left(t-1\right)=\left(\pi_{f}^{1}\left(t-1\right),\cdots,\pi_{f}^{Z_{f}}\left(t-1\right)\right).\label{eq:PreviousProbability}
\end{equation}
Here, we define ${\bf r}_{f}(t)=(r_{f}^{1}\left(t\right),\cdots,r_{f}^{m}\left(t\right)\cdots,r_{f}^{Z_{f}}\left(t\right))$
as a regret vector of determining flow-split vector for flow $f$.
The MBS selects the flow-split vector with highest regret in which
the mixed-strategy probability is given as

\vspace{-1em}
\begin{equation}
\pi_{f}^{m}\left(t\right)=\frac{\left[r_{f}^{m}\left(t\right)\right]^{+}}{\sum_{m'\in\mathcal{Z}_{f}}\left[r_{f}^{m'}\left(t\right)\right]^{+}}.\label{eq:highestregret}
\end{equation}
Let $\hat{{\bf r}}_{f}(t)=(\hat{r}_{f}^{1}\left(t\right),\cdots,\hat{r}_{f}^{m}\left(t\right)\cdots,\hat{r}_{f}^{Z_{f}}\left(t\right))$
be the estimated regret vector of flow $f$, we introduce the Boltzmann-Gibbs
(BG) distribution, $\boldsymbol{\beta}_{f}^{m}\left(\hat{{\bf r}}_{f}(t)\right)$\textcolor{blue}{{}
}\textcolor{black}{to capture the exploitation and exploration for
efficient learning, given by }
\begin{equation}
\begin{alignedat}{1}\boldsymbol{\beta}_{f}^{m}\left(\hat{{\bf r}}_{f}(t)\right)\:=\: & \underset{\boldsymbol{\pi}_{f}\in\Pi}{\mbox{argmax}}\sum_{m\in{\cal Z}_{f}}\left[\pi_{f}^{m}\left(t\right)\hat{r}_{f}^{m}\left(t\right)\right.\\
 & \qquad\quad\left.-\kappa_{f}\pi_{f}^{m}\left(t\right)\ln(\pi_{f}^{m}\left(t\right))\right],
\end{alignedat}
\label{eq:GibbsDistribution}
\end{equation}
where the trade-off factor $\kappa_{f}$ is used to balance between
exploration and exploitation \cite{bennislearning,2012Perlaza}.\textcolor{black}{{}
If $\kappa_{f}$ is small, the MBS selects ${\bf z}_{f}$ with highest
payoff. For $\kappa_{f}\rightarrow\infty$ all decisions have equal
probability.}

For a given set of $\hat{{\bf r}}_{f}(t)$ and $\kappa_{f}$, we solve
\eqref{eq:GibbsDistribution} to find the probability distribution
in which the solution determining the disjoint paths for each flow
$f$ is given as
\begin{equation}
\beta_{f}^{m}(\hat{{\bf r}}_{f}(t))=\frac{\exp\left(\frac{1}{\kappa_{f}}\left[\hat{r}_{f}^{m}\left(t\right)\right]^{+}\right)}{\sum\limits _{m'\in{\cal Z}_{f}}\exp\left(\frac{1}{\kappa_{f}}\left[\hat{r}_{f}^{m'}\left(t\right)\right]^{+}\right)}.\label{eq:GibbsSolution}
\end{equation}
We denote $\hat{u}\left(t\right)$ as the estimated utility of flow
$f$ at time instant $t$ with action ${\bf z}_{f}$, i.e, $\hat{{\bf u}}_{f}(t)=(\hat{u}_{f}^{1}\left(t\right),\cdots,\hat{u}_{f}^{m}\left(t\right)\cdots,\hat{u}_{f}^{Z_{f}}\left(t\right))$.
Upon receiving the feedback, $\tilde{u}_{f}(t)$ denotes the utility
observed by flow $f$, i.e., $\tilde{u}_{f}(t)=u_{f}(t-1)$. Finally,
we propose the learning mechanism at each time instant $t$ as follows.

\textbf{\textit{\textcolor{black}{Learning procedure}}}: The estimates
of the utility, regret, and probability distribution functions are
performed, and are updated for all actions as follows:\vspace{-1em}

\begin{equation}
\begin{cases}
\hat{u}_{f}^{m}\left(t\right)= & \hat{u}_{f}^{m}\left(t-1\right)+\xi_{f}(t)\mathbb{I}_{\{{\bf z}_{f}={\bf z}_{f}^{m}\}}\left(\tilde{u}_{f}(t)-\hat{u}_{f}^{m}\left(t-1\right)\right),\\
\hat{r}_{f}^{m}\left(t\right)= & \hat{r}_{f}^{m}\left(t-1\right)+\gamma_{f}(t)\left(\hat{u}_{f}^{m}(t)-\tilde{u}_{f}(t)-\hat{r}_{f}^{m}\left(t-1\right)\right),\\
\pi_{f}^{m}(t)= & \pi_{f}^{m}(t-1)+\iota_{f}(t)\left(\beta_{f}^{m}(\hat{{\bf r}}_{f}(t))-\pi_{f}^{m}(t-1)\right),
\end{cases}\label{eq:Probability}
\end{equation}
 Here, $\xi_{f}(t)$, $\gamma_{f}(t)$, and $\iota_{f}(t)$ are the
learning rates which are chosen to satisfy the convergence properties
\cite{2013backhaul}. Based on the probability distribution as per
\eqref{eq:Probability}, the MBS determines the flow-split vector
for each flow $f$ as defined in Section \ref{Pro-Form}. \textcolor{blue}{Note
that the learning-aided path selection is performed in a long-term
period to ensure that the paths do not suddenly change such that the
SCBSs have enough time to release traffic from the queues.}

\subsection{Rate Allocation}

Consider $R_{(i,j)}^{f}=\log(1+p_{(i,j)}^{f}|g_{(i,j)}({\bf h})|^{2})$
as the transmission rate, where the effective channel gain\footnote{The effective channel gain captures the path loss, channel variations,
and interference penalty (Here, the impact of interference is considered
small due to highly directional beamforming and high pathloss for
interfered signals at mmWave frequency band, and thus a multi-hop
directional transmission can be operated at dense mmWave networks). } for mmWave channels can be modeled as $g_{(i,j)}({\bf h})=\frac{\tilde{g}_{(i,j)}({\bf h})}{1+I^{\max}}$
\cite{2013beamforming,Vu_LB}. Here, $\tilde{g}_{(i,j)}({\bf h})$
and $I^{\max}$denote the normalized channel gain and the maximum
interference, respectively. Denoting the left hand side (LHS) of \eqref{eq:delayconst01}
and \eqref{eq:delayconst02} as $D_{i}^{f}$ for simplicity, the optimal
values of flow control ${\bf x}$ and transmit power ${\bf p}$ are
found by minimizing

\vspace{-1em}\begin{subequations}

{\small{}\label{RateAllocation}
\begin{alignat}{1}
\min_{{\bf x},{\bf p}|\boldsymbol{\pi}} & \quad\sum_{f=1}^{F}-Y_{f}x_{f}\label{eq:rateallocation}\\
\text{subject to} & \quad1+p_{(i,i_{f}^{({\rm o)}})}^{f}|g_{(i,i_{f}^{({\rm o)}})}|^{2}\geq e^{x_{f}},\forall f\in{\cal F},\,i=0,\label{eq:excontraint1}\\
 & \quad\frac{1+p_{(i,i_{f}^{({\rm o)}})}^{f}|g_{(i,i_{f}^{({\rm o)}})}|^{2}}{1+p_{(i_{f}^{({\rm I)}},i)}^{f}|g_{(i_{f}^{({\rm I)}},i)}|^{2}}\geq e^{D_{i}^{f}},\forall i\in{\cal N}_{i}^{\left(o\right)},f\in{\cal F},\label{eq:excontraint2}\\
 & \quad\sum_{f\in F}p_{(i,i_{f}^{({\rm o)}})}^{f}\leq P_{i}^{\text{max}},\forall i\in{\cal B},\forall f\in{\cal F}.\label{eq:powerconstraint}
\end{alignat}
}{\small \par}

\end{subequations}The constraint \eqref{eq:excontraint2} is non-convex,
but the LHS of \eqref{eq:excontraint2} is an affine-over-affine function,
which is jointly convex w.r.t the corresponding variables \cite{2004convex}.
In this regard, we introduce the slack variable $y$ to \eqref{eq:excontraint2}
and rewrite it as

\hspace{-1em}
\begin{alignat}{1}
\frac{2+p_{(i,i_{f}^{({\rm o)}})}^{f}|g_{(i,i_{f}^{({\rm o)}})}|^{2}}{2} & \geq\sqrt{y^{2}+\left(\frac{p_{(i,i_{f}^{({\rm o)}})}^{f}|g_{(i,i_{f}^{({\rm o)}})}|^{2}}{2}\right)^{2}},\label{eq:cone1}\\
\frac{y^{2}}{1+p_{(i_{f}^{({\rm I)}},i)}^{f}|g_{(i_{f}^{({\rm I)}},i)}|^{2}} & \geq e^{D_{i}^{f}}.\label{eq:cone2}
\end{alignat}
Here, the constraint \eqref{eq:cone1} holds a form of the second-order
cone inequalities \cite{2004convex,2001lectures}, while the LHS of
constraint \eqref{eq:cone2} is a quadratic-over-affine function which
is iteratively replaced by the first order to achieve a convex approximation
as follow

\vspace{-1em}
\begin{alignat}{1}
\frac{2yy^{(l)}}{1+p_{(i_{f}^{({\rm I)}},i)}^{f(l)}|g_{(i_{f}^{({\rm I)}},i)}|^{2}}-\frac{y^{(l)2}\left(1+p_{(i_{f}^{({\rm I)}},i)}^{f}|g_{(i_{f}^{({\rm I)}},i)}|^{2}\right)}{\left(1+p_{(i_{f}^{({\rm I)}},i)}^{f(l)}|g_{(i_{f}^{({\rm I)}},i)}|^{2}\right)^{2}}\geq e^{D_{i}^{f}}.\label{eq:cone3}
\end{alignat}
Here, the superscript $l$ denotes the $l$th iteration. Hence, we
iteratively solve the approximated convex problem of \eqref{RateAllocation}
as \textbf{Algorithm}~\ref{algRate1} in which the approximated problem
is given as

\begin{figure}
\includegraphics[width=1\columnwidth]{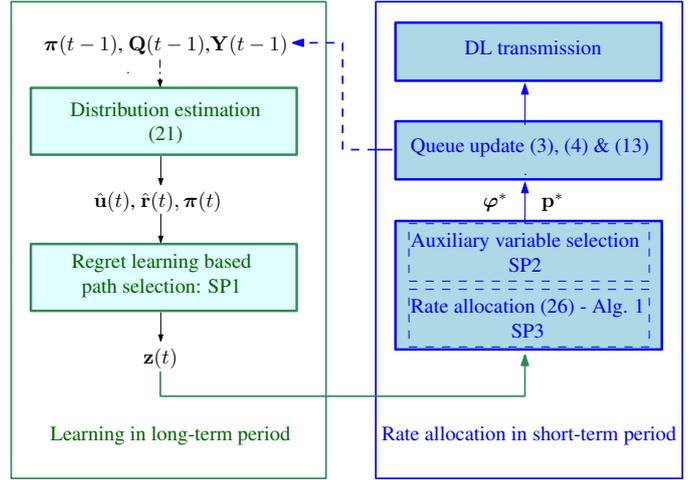}

\caption{Information flow diagram of the learning-aided path selection and
rate allocation.}
\label{Flowchart}
\end{figure}

\vspace{-1em}
\begin{eqnarray}
\min_{{\bf x},{\bf p}|\boldsymbol{\pi}} &  & \sum_{f=1}^{F}-Y_{f}x_{f}\label{Optimal-Rate}\\
\text{subject to} &  & \eqref{eq:powerconstraint},\,\eqref{ratebounds},\,\eqref{eq:excontraint1},\,\eqref{eq:cone1},\,\eqref{eq:cone3}.\nonumber
\end{eqnarray}
Finally, the information flow diagram of the learning-aided path selection
and rate allocation approach is shown in Fig. \ref{Flowchart}, where
the rate allocation is executed in a short-term period. Note that
the path selection and rate allocation are both done at the MBS, in
this work we assume that the information is shared among the base
stations by using the X2 interface.

\begin{algorithm}
\label{algRate}\begin{algorithmic}

\STATE Initialization: set $l=0$ and generate initial points $y^{(l)}$.

\REPEAT \STATE $\text{Solve}$~(\ref{Optimal-Rate}) with $y^{(l)}$
to get the optimal value $y^{(l)\star}$.

\STATE $\text{Update}$ $y^{(l+1)}:=y^{(l)\star}$; $l:=l+1$.

\UNTIL{\text{Convergence}}

\end{algorithmic} \caption{Iterative rate allocation}
\label{algRate1}
\end{algorithm}

Finally, the information flow diagram of the learning-aided path selection
and rate allocation approach is shown in Fig. \ref{Flowchart}, where
the rate allocation is executed in a short-term period.\textcolor{black}{{}
The reason why we chose an iterative method to solve the non-convex
optimization problem \eqref{RateAllocation} due to that in general
speaking, there is no fast and cost-efficient approach to solve \eqref{RateAllocation}.
Besides, finding the globally optimal solution for problem \eqref{RateAllocation}
via a brute-force approach entails a prohibitively high computational
complexity. Hence, we propose an iterative algorithm to obtain an
efficient suboptimal solution. }

\vspace{-0.8em}

\section{Numerical Results}

\label{Evaluation} We provide numerical results by assuming two flows
from the MBS to two UEs, while the number of available paths for each
flow is four \cite{key2011pathselection}. The MBS selects two routes
from four most popular routes\footnote{As studied in~\cite{key2011pathselection}, it suffices for a flow
to maintain at least two paths provided that it repeatedly selects
new paths at random and replaces if the latter provides higher throughput.}. Each route contains two relays, the total number of SCBSs is 8,
and the one-hop distance is varying from 50 to 100 meters. The maximum
transmit power of MBS and each SCBS are $43$ dBm and $30$ dBm, respectively.
The SCBS antenna gain is $5$ dBi and the number of antennas at each
BS is $N_{b}=8$. We assume that the traffic flow is divided equally
into two subflows, the arrival rate for each sub-flow is varying from
$2$ to $5$ Gbps. The path loss is modeled as a distance-based path
loss with the line-of-sight (LOS) model for urban environments at
$28$\,GHz with $1$ GHz of bandwidth~\cite{Vu_LB}. The maximum
delay requirement $\beta$ and the target reliability probability
$\epsilon$ are set to be $10~\text{ms}$ and $0.05$, respectively
\cite{vu2017ultra}. For the learning algorithm, the Boltzmann temperature
$\kappa_{f}$ is set to $5$, while the learning rates $\xi_{f}(t)$,
$\gamma_{f}(t)$, and $\iota_{f}(t)$ are set to $\frac{1}{\left(t+1\right)^{0.5}}$,
$\frac{1}{\left(t+1\right)^{0.55}}$, and $\frac{1}{\left(t+1\right)^{0.6}}$,
respectively \cite{bennislearning}.

\begin{figure}
\includegraphics[width=1\columnwidth]{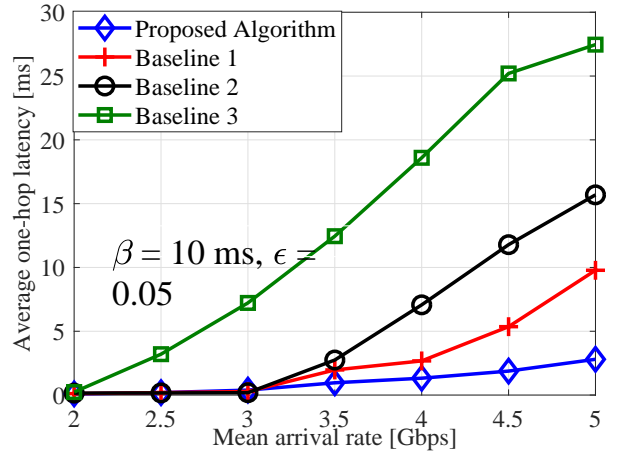}

\caption{Average one-hop delay versus mean arrival rates.}
\label{CCDF}
\end{figure}

Furthermore, we compare our proposed scheme with the following baselines:
\begin{itemize}
\item \textbf{\textit{Baseline}}\textbf{ 1 }considers a general NUM framework
\cite{neely2010S} with best path learning \cite{bennislearning}.
\item \textbf{\textit{Baseline}}\textbf{ 2 }considers a general NUM framework
\cite{neely2010S} and a random path section scheme, subject to $\eqref{eq:delayconst}$.
\item \textbf{\textit{Baseline}}\textbf{ 3} considers a general NUM framework
\cite{neely2010S} and a random path section scheme.
\item \textbf{\textit{Single hop}} scheme: The MBS delivers data to UEs
over one single hop at long distance in which the probability of LOS
communication is low, and blockage is taken into account.
\end{itemize}
In Fig.~\ref{CCDF}, we report the average one-hop delay\footnote{The average end-to-end delay can be defined as the sum of the average
one-hop delay of all hops.} versus the mean arrival rates $\bar{\mu}$. As we increase $\bar{\mu}$,
\textbf{\textit{baseline}} \textbf{3} violates the latency constraints,
whereas our proposed algorithm outperforms the other \textbf{\textit{baselines}}.
The reason behind this gain is that the delay requirement is satisfied
via the equivalent instantaneous rate by our proposed algorithm as
per \eqref{eq:delayconst01} and \eqref{eq:delayconst02}, while the
\textbf{\textit{baselines}} \textbf{1} and \textbf{3} use the traditional
utility-delay trade-off approach. Moreover, we apply the learning
path algorithm, which selects the path with high payoff and less congestion
resulting in small delay. The average one-hop delay of \textbf{\textit{baseline}}
\textbf{1} with learning outperforms \textbf{\textit{baselines}} \textbf{2}
and \textbf{3}, whereas our proposed scheme reduces latency by $50.64\%$,
$81.32\%$ and $92.9\%$ as compared to \textbf{\textit{baselines}}
\textbf{1}, \textbf{2}, and \textbf{3}, respectively, when $\lambda=4.5$\,Gbps.
When $\lambda=5$\,Gbps, the average delay of all \textbf{\textit{baselines}}
increases, violating the delay requirement of $10$\,ms, while our
proposed scheme is robust to the latency requirement. Moreover, for
throughput comparison, we observe that for $\lambda=4.5$\,Gbps,
our proposed algorithm is able to deliver $4.4874$ Gbps of average
network throughput per each subflow, while the \textbf{\textit{baselines}}
\textbf{1, 2,} and \textbf{3} deliver $4.4759$, $4.4682$, and $4.3866$
Gbps, respectively. Here, the \textbf{\textit{single hop}} scheme
only delivers $3.55$ Gbps due to the blockage, which resulting in
large delay.

\begin{figure}
\includegraphics[width=1\columnwidth]{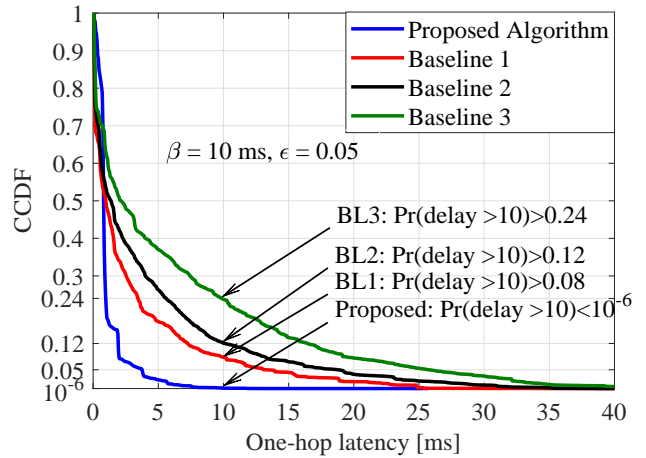}
\caption{CCDF of the one-hop latency, $\bar{\mu}=4.5$ Gbps.}
\label{avgDelay}
\end{figure}

In Fig.~\ref{avgDelay}, we report the tail distribution (complementary
cumulative distribution function (CCDF)) of latency to showcase how
often the system achieves a delay greater than the target delay levels.
In contrast to the average delay, the tail distribution is an important
metric to reflect the URLLC characteristic. For instance, at $\lambda=4.5$\,Gbps,
by imposing the probabilistic latency constraint, our proposed approach
ensures reliable communication with \textbf{better} guaranteed probability,
i.e, $\text{Pr}(\text{delay}>10\text{ms})<10^{-6}$. In contrast,
\textbf{\textit{baseline}} \textbf{1} with learning violates the latency
constraint with high probability, where $\text{Pr}(\text{delay}>10\text{ms})=0.08$
and $\text{Pr}(\text{delay}>25\text{ms})<10^{-6}$, while the performance
of \textbf{\textit{baselines}} \textbf{2} and \textbf{3} gets worse.

\section{Conclusion}

\label{Conclusion} In this paper, we have proposed a multi-hop scheduling
to support reliable communication by incorporating the probabilistic
latency constraint in 5G self-backhauled mmWave networks. In particular,
the problem is modeled as a network utility maximization subject to
a bounded latency constraint with a guaranteed probability, and queue
stability. We have proposed a dynamic approach, which adapts to channel
variations and system dynamics. We have leveraged stochastic optimization
to decouple the studied problem into path selection and rate allocation
sub-problems. Numerical results show that our proposed framework reduces
latency by $50.64\%$ and $92.9\%$ as compared to \textit{baselines}.

\vspace{-1em}

\section*{Acknowledgment}

The authors would like to thank Tekes, Nokia, Huawei, MediaTek, Keysight,
Bittium and Kyynel for project funding. The Academy of Finland funding
via the grant 307492 and the CARMA grants 294128 and 289611, the Nokia
Foundation, the Riitta and Jorma J. Takanen Foundation SR grant, the
Tauno Tönning Foundation, and the Finnish Technological Promotion
Foundation are also acknowledged.

\bibliographystyle{IEEEtran}
\bibliography{mmWave}

\end{document}